\documentclass[12pt,preprint]{aastex}
\usepackage{natbib}

\shortauthors{P\'erez-Gonz\'alez et al.}

\begin{document}

\title{LUMINOSITY AND STELLAR MASS FUNCTIONS OF LOCAL STAR-FORMING GALAXIES}

\author{Pablo G. P\'erez-Gonz\'alez, Jes\'us Gallego, Jaime Zamorano\altaffilmark{1}}

\author{Almudena Alonso-Herrero\altaffilmark{2}}

\author{Armando Gil de Paz\altaffilmark{3}}

\and 

\author{Alfonso Arag\'on-Salamanca\altaffilmark{4}}

\altaffiltext{1}{Departamento de Astrof\'{\i}sica, Facultad de F\'{\i}sicas, Universidad Complutense, E-28040 Madrid, Spain. E-mails: pag@astrax.fis.ucm.es, jgm@astrax.fis.ucm.es, jaz@astrax.fis.ucm.es}
\altaffiltext{2}{Steward Observatory, The University of Arizona, Tucson AZ 85721, USA. E-mail: aalonso@elgreco.as.arizona.edu}
\altaffiltext{3}{The Observatories of the Carnegie Institution of Washington, 813 Santa Barbara St., Pasadena, CA 91101, USA. E-mail: agpaz@ipac.caltech.edu}
\altaffiltext{4}{School of Physics and Astronomy, University of Nottingham, NG7 2RD, UK. E-mail: Alfonso.Aragon@nottingham.ac.uk}

\begin{abstract} 

We present the optical and near-infrared luminosity and mass functions of the
local star-forming galaxies in the {\it Universidad Complutense de Madrid}
(UCM) Survey. A bivariate method that explicitly deals with the H$\alpha$
selection of the survey is used when estimating these functions. Total stellar
masses have been calculated on a galaxy-by-galaxy basis taking into account
differences in star formation histories.  The main difference between the
luminosity distributions of the UCM sample and the  luminosity functions of the
local galaxy population is a lower normalization ($\phi^*$), indicating  a
lower global volume density of UCM galaxies.  The typical  near-infrared
luminosity ($L^*$) of local star-forming galaxies  is fainter than that of
normal galaxies. This is  a direct consequence of the lower stellar masses of
our objects. However, at optical wavelengths   ($B$ and $r$) the luminosity 
enhancement arising from the young stars leads to $M^*$ values that  are
similar to those of normal galaxies.  The fraction of the total optical and
near infrared  luminosity density in the local Universe associated with
star-forming galaxies is 10--20\%. Fitting the total stellar mass function
using a Schechter parametrization we obtain $\alpha=-1.15\pm0.15$,
$\log(\mathcal{M}^*)=10.82\pm0.17$~$\mathcal{M}_\sun$ and
$\log(\phi^*)=-3.04\pm0.20$~Mpc$^{-3}$.  This gives an integrated total stellar
mass density of $10^{7.83\pm0.07}$~$\mathcal{M}_\sun$~Mpc$^{-3}$ in local
star-forming galaxies ($\mathrm H_{0}=70$~km\,s$^{-1}$\,Mpc$^{-1}$,
$\Omega_{\mathrm M}=0.3$, $\Lambda=0.7$).  The volume-averaged burst strength
of the UCM galaxies is $b=0.04\pm0.01$, defined as the ratio of the mass
density of   stars formed in recent bursts (age$\,<10$~Myr)  to the total
stellar mass density in UCM galaxies.  Finally, we derive that, in the local
Universe, $(13\pm3)$\% of the total baryon mass density in the form of stars is
associated with star-forming galaxies.

\end{abstract}
\keywords{galaxies: luminosity function, mass function --- galaxies: 
starburst ---  infrared: galaxies}


\section{Introduction}

The star formation activity is a topic of growing interest as we extend the
study of the galaxy population to higher and higher redshifts. Part of this
interest is due to the fact that many  distant objects show an intense
starbursting activity \citep[see, e.g.,][ and references
therein]{1997ARA&A..35..389E,2000ARA&A..38..667F}.  One of the issues that is
being addressed is the characterization of the global properties of the galaxy
population as a function of lookback time. One of the most important of these
properties is the comoving star formation rate (SFR) density of the Universe
\citep[see, for example,][]{2000ApJ...544..641H,2001MNRAS.320..504S,
2001ApJ...549..745R,2002ApJ...570..492L}.  Another issue of
cosmological interest is the luminosity evolution of the galaxies
along the Hubble time. One way to study this evolution is through the
measurement of the number density of galaxies of a given luminosity as
a function of redshift, i.e., the time-evolution of the luminosity
function (LF).  A considerable amount of effort has been devoted to
the determination of the LF at a variety of redshifts and spectral
ranges \citep[among others,][]{1988MNRAS.232..431E,
1999ApJ...519....1S,2000MNRAS.312..557L,2001ApJ...560..566K,
2001MNRAS.326..255C,2002MNRAS.336..907N}.

Mass is even more fundamental than luminosity among the parameters
governing the formation and evolution of galaxies. Indeed, it may be
possible to distinguish between hierarchical and single collapse
models of galaxy formation using the mass of the observed galactic
structures and the history of their assembly as the differentiating
parameters
\citep[see][]{1998MNRAS.297..427A,1998ApJ...498..504B,1998ApJ...503..518F,
1999MNRAS.303..188K}.   Therefore, it is
extremely important to characterize the stellar and total masses of galaxies at
all redshifts and the number density of objects of a given mass, i.e., the
stellar mass function (SMF).

In this Letter, we have used the multi-wavelength dataset for the {\it
Universidad Complutense de Madrid} (UCM) Survey of emission-line
galaxies \citep[ELGs,][]{1994ApJS...95..387Z,1996ApJS..105..343Z} to
estimate the LFs and SMF of a complete sample of local star-forming
galaxies in the optical and near-infrared (NIR). Given that the UCM
Survey galaxies were selected by their active star-formation, our
results are directly comparable to the ones achieved for samples of
star-forming galaxies at high redshift, though one must be aware of
the differences on the selection techniques. Unless otherwise
indicated, throughout this Letter we use a cosmology with $\mathrm
H_{0}=70$~km\,s$^{-1}$\,Mpc$^{-1}$, $\Omega_{\mathrm M}=0.3$ and
$\Lambda=0.7$.


\section{The sample}

The UCM sample contains 191 local star-forming galaxies at redshifts lower than
0.045, with an average $\langle z\rangle  =0.026$. These objects were selected
from an objective-prism survey carried out with the Schmidt Telescope at Calar
Alto Observatory (Almer\'{\i}a, Spain) and centered at the wavelength of the
H$\alpha$ nebular emission \citep{1994ApJS...95..387Z,1996ApJS..105..343Z}.
Comparisons with other samples of star-forming galaxies selected using  a
variety of techniques  reveal that the UCM sample is representative of the
local population of star-forming galaxies 
\citep{1997ApJ...475..502G,2001AJ....121...66S,
1989ApJ...347..152S,1998MNRAS.300..303T}. Indeed,  the SFR density of the local
Universe derived from the UCM sample is reasonably  consistent with that
derived from other samples  \citep[see,
e.g.,][]{1998MNRAS.300..303T,2002MNRAS.330..621S}.

Optical spectroscopy is available in \citet{1996A&AS..120..323G}, and
photometry in the Gunn-$r$ and Johnson-$B$ filters can be  found in
\citet{1996A&AS..118....7V} and \citet{2000A&AS..141..409P}, respectively. 
NIR data ($J$ and $K_s$ bands) were presented in
\citet{2000MNRAS.316..357G} and \citet{2003MNRAS.338..508P}. An H$\alpha$ 
imaging study has also been carried out recently
\citep{2003ApJnotyet}. For a complete compendium of the spectroscopic 
and photometric properties of the sample, the reader may refer to
\citet[][and references therein]{2003MNRAS.338..508P} and 
http://www.ucm.es/info/Astrof/UCM\_Survey/survey.html.

In this Letter we present LFs in the 4 available optical and NIR  bands:
Johnson-$B$, Gunn-$r$, $J$ and $K_s$. Out of the original 191 galaxies, 15 were
classified as Active Galactic Nuclei by \citet{1996A&AS..120..323G} and have
been excluded. A further 11 galaxies have no data for  the line+continuum
magnitudes (see Section~\ref{method}). Thus, the sample used in this Letter is
composed of 165 objects.  The total numbers of objects with available
photometry in each band are: 165 in $B$, 142 in $r$, 143 in $J$ and 164 in
$K_s$.

The luminosities have been corrected for Galactic extinction using the maps of
\citet{1998ApJ...500..525S} and the extinction curve in
\citet{1989ApJ...345..245C}.  We have also applied k-corrections from 
\citet{1999A&A...351..869F} for $BJK_s$ and \citet{1995PASP..107..945F} for
Gunn $r$, taking into account the morphological types of the UCM galaxies
\citep{2001A&A...365..370P}. The k-corrections applied are small 
because of the low redshifts of our galaxies ($z<0.045$).

Total stellar masses for nearly all the UCM galaxies (156 objects) were
calculated in \citet{2003MNRAS.338..525P} using a complete set of stellar
population synthesis models to reproduce the broad-band and emission-line
spectral features in the optical and NIR. Taking into account different star
formation histories, the technique calculated mass-to-light ratios in the
$K_s$-band ($\mathcal{M}/L_{K_s}$) on a galaxy-by-galaxy basis. This Letter
uses the results on masses and burst strengths (ratio of the mass of the
stars younger than 10~Myr and the total stellar mass) which were calculated in
\citet{2003MNRAS.338..525P}, assuming an instantaneous burst of star formation
with a \citet{1955ApJ...121..161S} initial mass function
($\mathcal{M}_{\mathrm{low}}=0.1\,\mathcal{M}_\odot$ and
$\mathcal{M}_{\mathrm{up}}=100\,\mathcal{M}_\odot$) occurring on a `normal'
spiral galaxy. We used the stellar evolution synthesis library of Bruzual \&
Charlot (1999, private communication). The attenuation of the burst luminosity
was modelled with the recipe given in \citet{2000ApJ...539..718C}. The stellar
masses obtained using the alternative extinction recipes considered in 
\citet{2003MNRAS.338..525P} were statistically very similar. Thus, the
choice of extinction recipe has negligible effect on  the results presented
here.

\section{Estimating the luminosity and stellar mass functions}
\label{method}

The selection of the UCM Survey galaxies depends primarily on the
H$\alpha$ line flux and its contrast against the continuum
\citep{1995ApJ...455L...1G}. This contrast may be measured by a 
line+continuum magnitude $m_{L+C}$ \citep{1989ApJ...347..152S}. In the
present work we are concerned with the calculation of the luminosity
and mass functions of the sample in broad-band filters far away from
the passband were the selection took place. Therefore, a proper way of
estimating these functions is using a bivariate luminosity function
(BLF), $\phi(M_{L+C},M_n)$, which allows for sample selection in one
particular magnitude ($m_{L+C}$), 
and may be integrated over this magnitude to
obtain the LF in the required filter, $\phi(M_n)$. The method is
analogous to the one described in
\citet{2000MNRAS.312..557L}. It is an extension of the stepwise
maximum-likelihood technique presented in \citet{1988MNRAS.232..431E},
which properly handles the effects of not observing a small fraction
of the entire sample in the band where we are estimating the LF
\citep{1993MNRAS.260..285S}.

All the LF data points have been fitted to \citet{1976ApJ...203..297S}
functions taking the observational errors into account. The
uncertainties in the fitted parameters have been estimated considering
uncorrelated random variations of the LF points following a Gaussian
distribution with its width set by the errors of the data points.  We
calculate the errors in $\alpha$, $M^*$ and $\phi^*$ by repeating the
fit for each set of points with small random variations.

The bivariate method of estimating LFs has also been applied to the
calculation of the total stellar mass and burst mass functions of the
UCM sample, i.e., the volume density of local star-forming galaxies
and local starbursts as a function of mass. The use of this procedure
when estimating the SMF relies on principles similar to those of the
LF calculations. It uses a bivariate function
$\phi(M_{L+C},\mathcal{M}_*)$ defined as the number density of
galaxies of a given $M_{L+C}$ and $\mathcal{M}_*$. Note that the SMF
is often obtained by multiplying the $K_s$-band LF by a constant
$\mathcal{M}/L_{K_s}$.  However, several authors
\citep{2000ApJ...536L..77B,2001ApJ...550..212B,2001ApJ...559..620P,
2003MNRAS.338..525P} have suggested there could be sizeable
galaxy-to-galaxy variations in $\mathcal{M}/L_{K_s}$ because of
differences in star formation histories. Therefore, it seems safer to
estimate the SMF using individually-estimated $\mathcal{M}/L_{K_s}$
ratios for each galaxy, as we do here.

\section{Results}

\subsection{Luminosity functions}

In order to test the BLF method, we used it to re-calculate 
the H$\alpha$ LF of the
UCM sample. The results are almost identical to the ones achieved with
the $V/V_\mathrm{max}$ method
\citep{1968ApJ...151..393S,1973ApJ...186..433H} used by
\citet{1995ApJ...455L...1G} and 
\citet[][]{2003ApJnotyet}. This adds confidence to our results.



\begin{figure}
\begin{center}
\includegraphics[angle=-90,width=16cm]{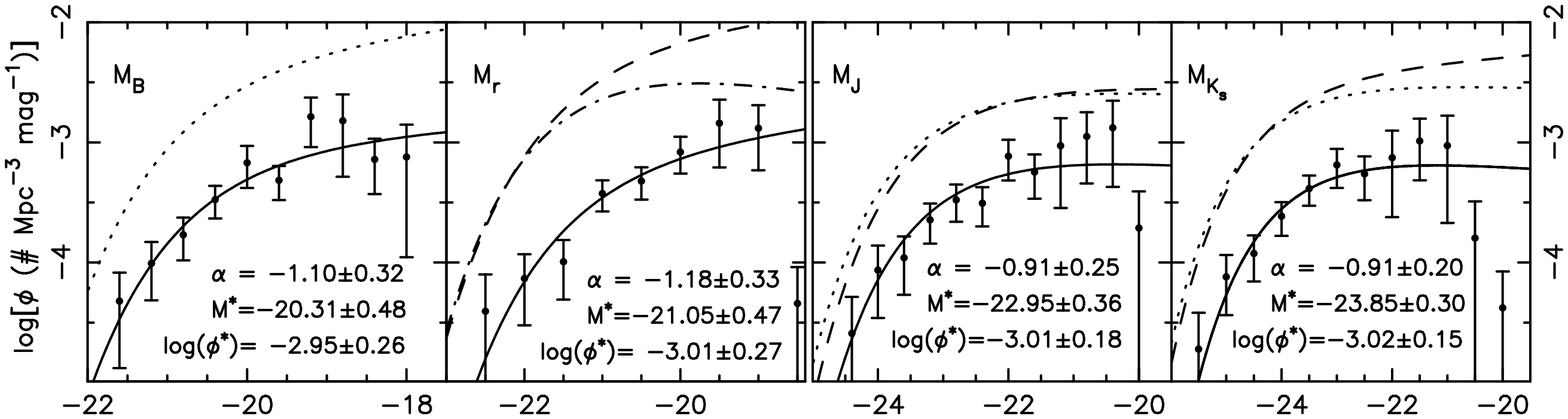}
\end{center}
\figcaption{\label{fig1}Luminosity functions for the UCM sample in the 
Johnson-$B$, Gunn-$r$, $J$ and $K_s$ bands. The Schechter fit parameters are
given inside each panel. The units of $\phi^*$ are Mpc$^{-3}$. The LFs of
several samples of `normal' galaxies  have also been plotted for comparison
(see text for details).}
\end{figure}

\placefigure{fig1}

Figure~\ref{fig1} shows the LFs of the UCM sample in $BrJK_s$. For comparison,
we also plot the LFs of the local galaxy population  obtained from different
surveys. The dotted line in the left panel of Figure~\ref{fig1} shows the
$B$-band LF of the 2dF Galaxy Redshift Survey
\citep[2dFGRS,][]{2002MNRAS.336..907N}. These galaxies have  redshifts up to
$\simeq0.2$, with  a modal  $z$ of $0.05$. The main different between the 
2dFGRS LF and the UCM LF is a lower  number density normalization
($\phi^*_\mathrm{UCM}/\phi^*_\mathrm{2dFGRS}=0.19$). This is not a surprise 
since not all the galaxies in the 2dFGRS emit in H$\alpha$. The $r$-band LF 
(second panel of Figure~\ref{fig1})  is compared to the LFs derived from the
Las Campanas Redshift Survey  \citep[LCRS; dash-dot line; $\langle
z\rangle\simeq0.1$; ][]{1996ApJ...464...60L},  and the Century Survey 
\citep[CS; dashed line; $z\le0.15$; ][]{1997AJ....114.2205G}\footnote{These LFs
have been transformed into the $r$-band using   $r-R_C=0.36$, typical of spiral
galaxies \citep{1995PASP..107..945F}. Note also  that these LFs have been
obtained using  different $\Omega_{\mathrm M}$ and $\Lambda$ values, but given
the moderate redshift of the samples we have not made any correction  in order
to match the cosmologies.}.  The comparison reveals  again a lower number
density of UCM galaxies    ($\phi^*_\mathrm{UCM}/\phi^*_\mathrm{LCRS}=0.15$ and
$\phi^*_\mathrm{UCM}/\phi^*_\mathrm{CS}=0.11$).   The $J$- and $K_s$-band LFs
are compared with those of the 2dFGRS \citep[dotted lines in the last two
panels of Figure~\ref{fig1}; ][] {2001MNRAS.326..255C}, and  the Two Micron All
Sky Survey \citep[2MASS; dashed lines; ][] {2000AJ....119.2498J}.  The $J$-band
2MASS LF comes from \citet{2001ApJ...557..117B},  with the global normalization
of \citet{2001MNRAS.326..255C}.  The $K_s$  2MASS LF was published by
\citet{2001ApJ...560..566K}. In both bands a normalization offset is observed
again: $\phi^*_\mathrm{UCM}/\phi^*_\mathrm{2dFGRS}=0.27$ in $J$; 
$\phi^*_\mathrm{UCM}/\phi^*_\mathrm{2dFGRS}=0.26$ and 
$\phi^*_\mathrm{UCM}/\phi^*_\mathrm{2MASS}=0.24$ in $K_s$.


The shapes of the luminosity distributions of the UCM sample are not
very different from those of the global population of galaxies in the
local Universe.  However, since the UCM galaxies have typically higher
star formation activity than ``normal'' quiescent spiral galaxies
\citep{2003MNRAS.338..525P}, a lower number density is found.  The
ratios between the {\it integrated} luminosity densities for the local
star-forming objects in the UCM Survey and for the global population
of galaxies are $0.15\pm0.05$, $0.11\pm0.04$, $0.23\pm0.04$,
$0.18\pm0.03$ in $B$, $r$, $J$, and $K_s$. Note that the local galaxy
population LFs in the $B$, $J$ and $K_s$~bands have been derived in a
homogeneous way from the 2dFGRS.  Remarkably, these luminosity density
ratios are comparable at all wavelengths and very similar to the ratio
of the number of enhanced star-forming galaxies to the total
\citep{1996ApJS..105..343Z}. The luminosity density ratios do not
depend strongly on luminosity. Moreover, these ratios are a factor of
$\sim2$ higher than the nearby galaxy-pair fraction, estimated to be a
6--10\% by
\citet{1992A&AS...94..553K}. This might suggest that a significant fraction  of
the star-formation activity in the local Universe is driven by minor
mergers, where the smaller in-falling object is not always detectable.

In the NIR, the UCM galaxies yield fainter $M^*$ values (by $0.4$~mag) than the
`normal' galaxy samples.   This is directly related to the lower typical 
stellar masses found for the local star-forming objects
\citep{2003MNRAS.338..525P}:  the $J$ and $K_s$ luminosities are strongly
correlated with the total stellar mass, while being less influenced by  recent
star formation. However, the enhanced star formation is responsible for the UCM
Survey presenting $M^*$ values similar to those of normal galaxies at optical
wavelengths despite having smaller stellar masses.

\subsection{Mass functions}

\begin{figure}
\begin{center}
\includegraphics[angle=-90,width=8cm]{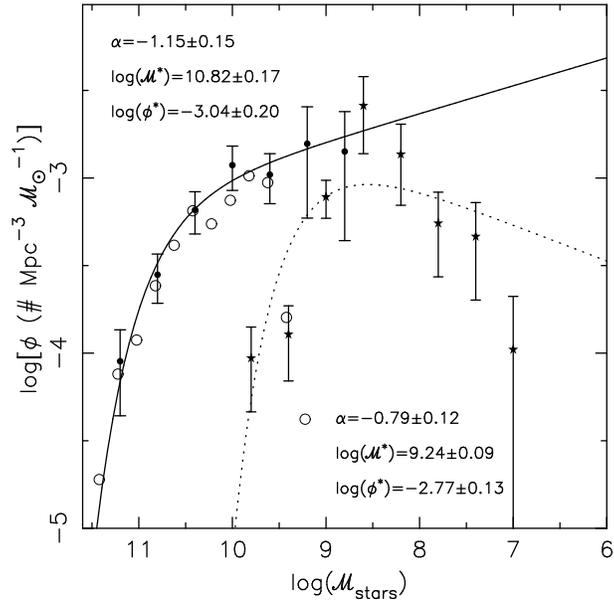}
\end{center}
\figcaption{\label{fig2}Total stellar mass function  (filled circles and solid
line)  and burst mass function (stars and dotted line) for the UCM sample. Open
symbols show the mass function calculated by multiplying the $K_s$-band LF by a
constant $\mathcal{M}/L_{K_s}=0.78$. The Schechter fit parameters are given for
the total masses (upper-left corner) and burst masses (lower-right corner).
Masses are in solar units and $\phi^*$ in Mpc$^{-3}$.}
\end{figure}

Figure~\ref{fig2} shows the total stellar mass function  (filled circles) and
burst mass function (stars) for the UCM galaxies calculated as described  in
Section~3. Schechter function  fits (solid and dotted lines respectively) are
also shown.   In addition, we have  calculated the SMF multiplying the
$K_s$-band LF by a constant $\mathcal{M}/L_{K_s}=0.78$ (in solar units), the
average mass-to-light ratio used in \citet{2003MNRAS.338..525P} for the evolved
stellar populations within the UCM galaxies (open circles).  At high masses,
this simple estimate of the SMF  coincides with the one
obtained using individual $\mathcal{M}/L_{K_s}$ values.  However, large
differences appear for low mass galaxies: these objects harbor the most intense
bursts \citep{2003MNRAS.338..525P} which have a major influence in the
$\mathcal{M}/L_{K_s}$ ratios. The errors introduced by using a constant
$\mathcal{M}/L_{K_s}$ are expected to be much larger at high-$z$ where the
burst strengths are higher.
 
A Schechter function seems to provide a reasonable fit to the total SMF.  The
typical stellar mass ($\mathcal{M}_*$) of  the local star-forming galaxies in
the UCM sample is a factor $\sim2$ smaller than that of the local galaxy
population   \citep{2001MNRAS.326..255C}. This is directly related to the
fainter NIR $M^*$ found for the UCM galaxies.   The faint end slope
($\alpha$)  of our SMF and the one published in \citet{2001MNRAS.326..255C} are
virtually identical. 


\placefigure{fig2}


We have integrated the total SMF in Figure~\ref{fig2} to obtain the
total stellar mass density in local star-forming galaxies,
$\rho_*^\mathrm{ELG}(z=0)=10^{7.83\pm0.07}$~$\mathcal{M}_\sun$~Mpc$^{-3}$. This
corresponds to $(13\pm3)$\% of the baryon mass density in the form of
stars and their remnants estimated by \citet{1992MNRAS.258P..14P},
\citet{1998ApJ...503..518F} and \citet{2001MNRAS.326..255C} for the local Universe.

Integrating the burst SMF allows us to calculate the volume-averaged
burst strength $b$ of these star-forming galaxies, defined as the
ratio of the burst mass density to the total stellar mass density. We
find $b(z=0)=0.04\pm0.01$.  This means that $\sim4$\% of the total
stellar mass in local star-forming galaxies corresponds to stars
younger than $\simeq10$~Myr
\citep{2003MNRAS.338..525P}.  Note that the galaxies within the UCM Survey have
$EW(\mathrm H\alpha)>20$~\AA\, \citep{1995ApJ...455L...1G}, so this value must
be considered as a lower limit to the relative fraction of young stars in our
local Universe.


It is interesting to compare our measure  of $\rho_*^\mathrm{ELG}(z=0)$ with
high redshift estimates. Our value is 1.6 times lower than the comoving stellar
mass density in spiral galaxies at $0.4<z<1$ \citep{2000ApJ...536L..77B}.
Also,  when  compared with the $z>0$ galaxy samples  presented in 
\citet{astro-ph/0212242}, our results point towards a decline in
$\rho_*^\mathrm{ELG}$ from $z\sim1$ to $z=0$.  This is  similar to the
behavior observed for the SFR density. One should be cautious when making
comparisons with  $z>1$,  since the high-$z$ samples are dominated by intense
star-formation and may present selection biases at least as important as the
ones in our sample, leading to an underestimate of the global $\rho_*$.


\section{Summary and conclusions}

In this {\it Letter\/} we have presented the luminosity and mass functions of
the local star-forming galaxies in the UCM Survey. Given that the UCM galaxies
were selected by their active star formation, our results are directly
comparable to those of high redshift star-forming galaxies. In estimating 
these functions we have accounted for the H$\alpha$ selection of our survey and
calculated stellar masses on a galaxy-by-galaxy basis.

The main results of our study are: (1) The luminosity functions of the
UCM Survey galaxies in the optical and near-infrared have lower
normalizations than the LFs of the global population of galaxies,
indicating a lower number density. This is not surprising since the
UCM galaxies have significantly higher star formation activity than
`normal' quiescent galaxies, and galaxies without star formation are
not present in the sample. (2) In the NIR, UCM galaxies have fainter
characteristic magnitudes ($M^*$) than `normal' galaxies. This is a
direct consequence of the lower typical stellar masses of our
objects. However, the optical $M^*$ values of the UCM galaxies are
similar to those of `normal' galaxies, suggesting the enhanced star
formation has boosted up their optical luminosities.  (3) The fraction
of the total luminosity density in the local Universe associated with
star-forming galaxies is $0.15\pm0.05$, $0.11\pm0.04$, $0.23\pm0.04$,
$0.18\pm0.03$ in $B$, $r$, $J$, and $K_s$. Remarkably, these luminosity
fractions are comparable at all wavelengths and do not depend strongly
on luminosity. This may be due to selection effects shared by all
magnitude (and surface brightness) limited surveys. (4) If we fit the
total stellar mass function using a Schechter parameterization, we
obtain $\alpha=-1.15\pm0.15$,
$\log(\mathcal{M}^*)=10.82\pm0.17$~$\mathcal{M}_\sun$ and
$\log(\phi^*)=-3.04\pm0.20$~Mpc$^{-3}$. This gives an integrated total
stellar mass density in star-forming galaxies in the local Universe of
$10^{7.83\pm0.07}$~$\mathcal{M}_\sun$~Mpc$^{-3}$.  (5) The
volume-averaged burst strength of the UCM galaxies is $b=0.04\pm0.01$,
defined as the ratio of the mass density of stars formed in recent
bursts (age$\,<10$~Myr) to the total stellar mass density in UCM
galaxies.  (6) We estimate that $(13\pm3)$\% of the total baryon mass
density in the form of stars is associated with star-forming galaxies
in the local Universe.

\acknowledgments

PGPG wishes to acknowledge the Spanish Ministry of Education and
Culture for the reception of a {\it Formaci\'on de Profesorado
Universitario} fellowship. AAS acknowledges generous financial support
from the Royal Society. AAH has been supported by the National
Aeronautics and Space Administration grant NAG 5-3042 through the
University of Arizona. AGdP acknowledges financial support from NASA
through a Long Term Space Astrophysics grant to B.F.\ Madore. We are
grateful to the anonymous referee for his helpful comments and
suggestions. Valuable discussion with Casey Papovich is
acknowledged. The present work was supported by the Spanish Programa
Nacional de Astronom\'{\i}a y Astrof\'{\i}sica under grant
AYA2000-1790.

\bibliographystyle{apj}
\bibliography{referencias}

\end{document}